
\magnification=1200
\catcode `@=11

\hsize 16truecm
\vsize 24truecm
\null
\rightline{CPT-93/P. 2883}
\rightline{UAB--FT--310}
\vskip 2truecm
\centerline{\bf CONSTITUENT QUARK COUPLINGS}
\centerline{\bf AND}
\centerline{\bf QCD IN THE LARGE $N_c$ LIMIT}
\vskip 1truecm
\centerline{{\bf Santiago PERIS}\footnote{$^a$}{E-mail: Peris @ IFAE.ES}
\footnote{$^\star$}{Address after September 1993 : CERN, Theory
Division}}
\centerline{Grup de Fisica Teorica i Institut de Fisica d'Altes Energies}
\centerline{Universitat Autonoma de Barcelona}
\centerline{08193 Bellaterra (Barcelona), Spain}
\bigskip
\centerline{\bf and}
\bigskip
\centerline{\bf Eduardo de RAFAEL}
\centerline{Centre de Physique Th\'eorique}
\centerline{CNRS - Luminy, Case 907}
\centerline{13288 Marseille Cedex 9, France}

\vskip 2truecm

\centerline{\bf Abstract}
\bigskip
It is shown that in the constituent quark model of Georgi-Manohar,
the dispersion relation that leads to the
Adler-Weisberger sum rule for the axial
vector coupling $g_A$ requires a subtraction constant.
This fact explains the discrepancy between the
results of different recent estimates of the $1/N_c$ corrections to
Weinberg's large $N_c$ result $g_A=1$, where $N_c$ is the number of QCD
colors.
We also discuss a possible scenario which shows that $g_A=1$ might not be a
necessary consequence of QCD in the large $N_c$ limit.

\vfill\eject

\baselineskip 18pt
The relationship between phenomenological quark models, like the De
R\'ujula-Georgi-Glashow model
[1], and quantum chromodynamics remains still an intriguing question
in hadron physics. A possible
scenario suggested by Georgi and Manohar [2], which we shall adopt
throughout this work, assumes that in the intermediate energy region
between
the scale at which the chiral $SU(3)$ flavour symmetry is spontaneously
broken ($\Lambda_{\chi}\simeq
1$GeV) and the confinement scale ($\Lambda_{\overline{MS}}\simeq 200$
MeV), QCD may be formulated in
terms of an effective field theory of constituent chiral quarks
interacting with the Goldstone modes
associated with the spontaneous chiral symmetry breaking; perhaps, with
 the inclusion of long distance
gluon interactions as well. The electroweak couplings of the constituent
 chiral     quarks may then be
calculable; at least in some appropriate approximation of QCD, like
for example     the large $N_c$
limit [3] where the number of colours $N_c$ is taken to infinity, with
the coupling $\alpha_sN_c$
held fixed. To leading order in the $1/N_c$-expansion; and assuming
that some specific pion-quark
and photon-quark scattering amplitudes obey unsubtracted dispersion
relations, Weinberg has shown
that constituent chiral quarks have an axial vector coupling $g_A=1$
 and no anomalous magnetic   moment [4].

Estimates of the corrections of order $1/N_c$ to the result $g_A=1$,
within the    framework of a
Gell-Mann-L\'evy like linear sigma model [5] for constituent quarks, have
been made in refs.[6]; and,
using the non-linear sigma model of Georgi and Manohar [2], in refs.[7].
 Although both estimates find
that numerically $g_A$ becomes smaller than one, i.e. a correction in the
right direction,  the form of the two results is surprisingly
different. While the linear sigma model calculations [6] give a
 correction:
 $$g^2_A=1-2{M^2_Q\over 16\pi^2f^2_{\pi}}\left(\log{M^2_\sigma\over
M^2_Q}+O(1)\right)+O\left[N_c\left({M^2_Q\over 16\pi^2f^2_{\pi}}\right)^2
\right]\eqno (1)$$
which grows logarithmically in the limit where
$M_{\sigma}$ becomes large with respect to the constituent quark mass
$M_Q(M_Q\simeq 300 MeV)$ ; the
calculations of refs.[7], using the analog of the Adler-Weisberger
sum rule [8]    for constituent
quarks, find (when normalized to the same definition of $f_{\pi}$,
$f_{\pi}\simeq 93$MeV)
$$g^2_A=1-2{M^2_Q\over
16\pi^2f^2_{\pi}}+O\left[N_c\left({M^2_Q\over16\pi^2f^2_{\pi}}\right)^2
\right]\ ;\eqno (2)$$
i.e. a result without a logarithmic term at order $\displaystyle{M^2_Q
\over16\pi    ^2f^2_{\pi}}$. To
lowest order in the chiral expansion, and for the first term in an
expansion in    powers of
$\displaystyle{M^2_Q\over16\pi^2f^2_{\pi}}$, one would expect the
results of the       two models to agree
in the limit where $M_{\sigma}\to \infty$. The main point of this
letter is a clarification about the
origin of these two different results. However, as we shall later
 discuss,   the explanation of
this discrepancy also raises the question of whether or not $g_A=1$
in the  large $N_c$ limit.

We shall first briefly review the main steps in the derivation of the
Adler-Weisberger sum rule for constituent quarks, which is the
 calculational framework used in
refs.[6b] and [7].

The relevant physical quantity is the forward elastic amplitude for
on-shell pion-quark scattering
($\nu =p.q$, with $p$ the energy-momentum of the quark and $q$ the
energy-momentum of the pion; $i$    and $j$ are isospin pion indices):
$$T^{ij}(\nu)=\delta^{ij}T^{(+)}(\nu)+{1\over 2}[\tau^i,\tau^j]{\nu\over
f^2_{\pi}}T^{(-)}(\nu)\ .\eqno (3)$$
As in the case of pion-nucleon scattering, the isospin odd amplitude
$T^{(-)}(\nu)$ is assumed to
satisfy an unsubtracted dispersion relation in the $\nu$-variable.
The optical theorem relates the
absorptive part of this amplitude to the difference of the total
$\pi^-$--up quark and $\pi^+$--up quark
cross sections $\sigma^{(-)}$ and $\sigma^{(+)}$. From the dispersion
relation,    it then follows that
$$\lim_{\nu\to 0}ReT^{(-)}(\nu)={2f^2_{\pi}\over\pi}\int_{0}^{\infty}
{d\nu\over
\nu}\left[\sigma^{(-)}(\nu)-\sigma^{(+)}(\nu)\right]\ .\eqno (4)$$
The low energy theorems of current algebra relate the left hand side
in eq.(4) to $g^2_A$, with the   result
$$\lim_{\nu\to 0}ReT^{(-)}(\nu)=1-g^2_A\ .\eqno (5)$$
In the framework of the constituent chiral quark model of Georgi and
 Manohar [2], this is the result
which follows from the simple calculation of the tree diagrams shown in
Fig. 1 in the chiral limit
$(m_{\pi}\to 0)$. (The seagull graph contributes the factor 1; each
 one of the other graphs a factor
$-{1\over 2}g^2_A$.) Combining eqs.(4) and (5) results in an
Adler-Weisberger like sum rule for constituent quarks.

In order to further investigate the issue in question, we propose to do
 a one-loop calculation of
the real part of the $T^{(-)}$-amplitude in eq.(3). Obviously, we shall
 calculate
$ReT^{(-)}(0)$ in the chiral limit, and within the same model which has
been used in refs.[7] to
calculate $\sigma^{(-)}(\nu)$ and $\sigma^{(+)}(\nu)$; i.e. the chiral
 quark model of Georgi and  Manohar [2] with $g_A=1$:
$${\cal L}_{GM}=i\overline Q\gamma^{\mu}\left(\partial_{\mu}+
\Gamma_{\mu}-{i\over
2}\gamma_5\xi_{\mu}\right)Q-M_Q\overline QQ+{1\over 4}f^2_{\pi}tr
\partial_{\mu}{\cal
U}\partial^{\mu}{\cal U}^{\dagger}\ ,\eqno (6a)$$
where
$$\Gamma_{\mu}={1\over
2}\left(\xi^{\dagger}\partial_{\mu}\xi+\xi\partial_{\mu}\xi^{\dagger}
\right) \quad\hbox{and}\quad
\xi_{\mu}=i\left(\xi^{\dagger}\partial_{\mu}\xi-\xi\partial_{\mu}\xi^
{\dagger}  \right)\ .\eqno (6b,c)$$
The full set of relevant one-loop Feynman diagrams is shown in Fig. 2.
The contribution from the sum
of diagrams (1) to (9) is already known from refs.[7]. The result
 must be the same as the one
obtained from the calculation of the absorptive parts of these diagrams,
 and then using an
unsubtracted dispersion relation; i.e. the Adler-Weisberger relation.
Therefore,
$$\left.ReT^{(-)}(0)\right|_{(1),(2),...,(9)}=2{M^2_Q\over16\pi^2f^2
_{\pi}}\ .\eqno (7)$$

Diagrams (10) and (11) were not considered in refs.[7]. They are however
 of the       same order in the
$1/N_c$-expansion as the other diagrams (1) to (9) in Fig. 2. The
reason why they      were not considered
in the calculation of refs.[7] is that, contrary to the other diagrams
in Fig. 2, diagrams
(10) and (11) have no discontinuity in a $\pi$-quark intermediate state;
 and therefore they do not
contribute to $\sigma^{(-)}$ and $\sigma^{(+)}$ at the order we are
considering    in the $1/N_c$-expansion.
They contribute however to $ReT^{(-)}(0)$. The local four-pion vertex
in these
diagrams is given by the pion field expansion of the lowest order
non-linear sigma model
two-derivative term in eq.(6a):
$${1\over 4}f^2_{\pi}tr\partial_{\mu}{\cal U}^{\dagger}\partial^{\mu}
{\cal U}={1  \over
2}\left(\partial_{\mu}\vec{\pi}\right)^2+{1\over2f^2_{\pi}}
\left(\vec{\pi}\cdot\partial_{\mu}\vec{\pi}\right)
\left(\vec{\pi}\cdot\partial^{\mu}\vec{\pi}\right)+\dots\ .\eqno (8)$$
The contribution to $ReT^{(-)}(0)$ from the sum of diagrams (10) and
(11) in Fig. 2 is
logarithmically divergent in the ultraviolet. The coefficient of the
 logarithm however can be
calculated unambiguously, with the result:
$$\left.ReT^{(-)}(0)\right|_{(10),(11)}=2{M^2_Q\over 16\pi^2f^2_{\pi}}
\left(\log    {\Lambda^2\over
M^2_Q} +Const.\right)\ .\eqno (9)$$
Inserting the total result for $ReT^{(-)}(0)$ obtained from the sum of
 eqs.(7) and (9), in
eq.(5), gives a determination of $g^2_A$ which coincides with the one
in eq.(1),     obtained in the
linear sigma model calculation of refs.[6], in the limit of a large
sigma mass i.e. $M_{\sigma}=\Lambda$.

{}From the results of the calculation discussed above, we see that the
amplitude
$T^{(-)}$, in the chiral quark model of Georgi and Manohar, does not
 obey an unsubtracted
dispersion relation to order $1/N_c$: the real part of $T^{(-)}$ from
diagrams (10) and (11) in Fig. 2 is not zero,
and cannot be obtained from a dispersive integral.

We should point out that
the fact that the logarithmic contribution in eq. (9) appears as a
subtraction in the Adler-Weisberger relation is not the
relevant point here, since this merely reflects the bad high-energy
behavior of the effective lagrangian that has been used. What is
important  is that this contribution is nonzero.
Notice that the effective lagrangian of eqs. (6) has $g_A=1$ but it is
otherwise universal, i.e. {\it any} model containing constituent quarks
 and    pions is described
by this lagrangian at low energies and therefore produces the same
logarithmic
contribution even though, in general, it may not appear as a subtraction
in the dispersion relation if the model  has ``good"
high-energy behavior.
After all, there are good reasons to believe that the Adler-Weisberger
sum
rule is unsubtracted [4].Therefore, although this logarithmic contribution
starts as a
subtraction in the effective theory at low energies,
it will eventually have to evolve, as the energy grows,
into an
ordinary contribution of the underlying theory to the dispersive
integral in eq. (4).
This can be exemplified with the linear sigma
model as a toy model. In perturbation theory, this model does not
require subtractions in the Adler-
Weisberger relation. However, at the one-loop level, it yields exactly the
same logarithmic term (with $M_{\sigma}$ instead of $\Lambda$) in
$g_A$, but through the integral in eq. (8). This is depicted in Fig. 3; the
calculation was done in ref.[6b] and yields eq.(1). This concludes our
discussion of the $1/N_c$ corrections to the large $N_c$ result $g_A=1$.

Let us now come back to the question we were referring to at the beginning
as to whether or not $g_A=1$ in the large $N_c$ limit.

It has been recently shown [9], that the constituent chiral quark model
 of Georgi and Manohar can be
viewed as a particular case of an extended Nambu Jona-Lasinio model [10]
 with four-quark interaction
couplings ($a$ and $b$ are $u,d,s$ flavour indices and colour
summation within each quark bilinear
bracket is implicit; $q_{L,R}={1\over 2} (1\mp \gamma_5) q$):
$${\cal L}_{S,P}={8\pi^2G_s(\Lambda_{\chi})\over N_c\Lambda^2_{\chi}}
\sum_{a,b}^   {}\left(\overline q^a_Rq^b_L\right)\left(\overline
q^b_Lq^a_R\right)\eqno (10a)$$  and
$${\cal L}_{V,A}=-{8\pi^2G_v(\Lambda_{\chi})\over
N_c\Lambda^2_{\chi}}\sum_{a,b}^{}\left[\left(\overline q^a_L\gamma^{\mu}
q^b_L\right)\left(\overline
q^b_L\gamma_{\mu}q^a_L\right)+L\leftrightarrow R\right]\ .\eqno (10b)$$
Notice that the operator of eq.(10b) contains the vector--vector as
 well as   the axial-vector--axial-vector combinations.
The scenario suggested in ref.[9] assumes that, at intermediate energies
 below or of the order of
the spontaneous chiral symmetry breaking scale $\Lambda_{\chi}$, these
 are the leading operators of
higher dimension which due to the growing of their couplings $G_s$ and
 $G_v$ as    the ultraviolet
cut-off approaches its critical value from above become relevant.
In QCD, and with the factor
$N_c^{-1}$ pulled out, both couplings $G_s$ and $G_v$ are $O(1)$
in the large $N_c$ limit. As is
well known in the Nambu Jona-Lasinio model [10], the
${\cal L}_{S,P}$-operator,     for values of
$G_s>1$, is at the origin of the spontaneous chiral symmetry breaking.
It is this operator which
generates a constituent chiral quark mass term (${\cal U}$ is a unitary
$3\times     3$ matrix which
collects the Goldstone field modes):
$$-M_Q\left(\overline q_L{\cal U}^{\dagger}q_R+\overline q_R{\cal U}
q_L\right)=-       M_Q\overline QQ\
,\eqno (11)$$ like the one which appears in the Georgi-Manohar model
[2]; as well as in the effective
action approach of ref.[11].

As discussed in ref.[9], the ${\cal L}_{V,A}$-operator leads to
an effective axial coupling of the
constituent quarks with the Goldstone modes $\left(\xi_{\mu}=i\xi^{
\dagger}\partial_{\mu}{\cal
U}\xi^{\dagger}\ ;\ \xi\xi={\cal U}\right)$:
$${i\over 2}g_A\overline Q\gamma^{\mu}\gamma_5\xi_{\mu}Q\ ,\eqno (12)$$
with
$$g_A={1\over 1+4G_v{M^2_Q\over \Lambda^2_{\chi}}
\int_{M^2_Q/\Lambda^2_{\chi}}^{\infty}{dz\over
z}e^{-z}}\eqno (13)$$
to leading order in the $1/N_c$-expansion. For $G_v\not = 0$,$M_Q\not =
0$   and finite $\Lambda_{\chi}$ this result
implies $g_A\not = 1$. In
terms of Feynman diagrams it can be understood as the infinite sum of
 constituent quark bubbles
shown in Fig. 4a., where the cross at the end represents the pion field.
 These are the diagrams
generated by the $G_v$-four fermion coupling to leading order in the
$1/N_c$-expansion. The quark
propagators in Fig. 4a, are constituent quark propagators, solution of
the Schwinger-Dyson equation in
the leading large $N_c$-approximation, which diagrammatically is
represented in Fig. 4b. In terms of mesons fields, Fig. 4a is nothing but a
mixing term between the pion and the axial vector.
Although the
result in eq.(13) certainly cannot be claimed to be the exact QCD answer
in the large $N_c$ limit, it is
nevertheless distressing to find a model of large $N_c$ interactions
which does not lead to a $g_A=1$
result. Can one safely claim that this is just the shortcoming of the
 model
and, consequently, disregard it ? or is the model pointing toward the
possibility that $g_A=1$ might not be a necessary consequence of QCD
in the large $N_c$ limit? .

It should be stressed that both four-fermion operators in eqs. (10) are
natural at scales $\mu < \Lambda_{\chi}$ since they are supposed to
describe effective interactions in the intermediate region
$\Lambda_{QCD} <
\mu < \Lambda_{\chi}$; and they are not forbidden by any symmetry of the
QCD lagrangian. In fact they can be viewed as a Fierz-reordered
 version of
QCD color quark current interactions:
$$\eqalign{\sum_A \bar q\gamma^{\mu} {\lambda^A \over 2}q
&\bar q\gamma_{\mu} {\lambda^A \over 2}q  = \cr
& 2\sum_{a,b} (\bar q_R^a q_L^b)(\bar q_L^b q_R^a) -
{1\over 2} \sum_{a,b}[ (\bar q_L^a \gamma^{\mu} q_L^b)
(\bar q_L^b \gamma_{\mu} q_L^a) + (L\to R)]
+ {\cal O}({1\over N_c})\ .\cr}\eqno (14)$$
(The notation is as in eqs. (10)).
The four-quark operators in eqs. (10) are expected to become
irrelevant at short
distances; i.e. at scales $\mu > \Lambda_{\chi}$ where $G_s < 1$ and no
solution, other than $M_Q=0$, exists for the gap equation. Therefore, they
do not necessarily contradict the expected convergence properties of the Adler-
Weisberger sum rule in the high-energy regime. They do, however, play an
important role at long distances, and in particular they lead to the
result
$g_A\not =1$ in the large $N_c$ limit. This result appears as the
interplay
between spontaneous chiral symmetry breaking $(M_Q\not =0)$ and the
effect
of mixing between the pion field and the axial-vector field
($G_V\not = 0 $
and finite $\Lambda_{\chi}$). Both are expected features of QCD. The
summation of an {\it infinite} class of large $N_c$ QCD diagrams is however
crucial to implement spontaneous chiral symmetry breaking
($M_Q\not =0$)[12] and the
appearance of a scale like $\Lambda_{\chi}$.
It is only after this infinite summation that operators like those in eqs.
(10) may appear, giving rise to $g_A\not= 1$, and eluding the large
$N_c$ arguments based on a finite subset of diagrams that led to the result
$g_A=1$ [4].


With regards to the anomalous magnetic moment of the constituent quark,
the situation seems different.
In order to get a correction to $g-2$ of $O(1)$ in the
$1/N_c$-expansion, one would need an operator
like the one in eq.(10b) with $\sigma^{\mu\nu}$ istead of $\gamma^{\mu}$.
For large $N_c$, this operator cannot arise
in QCD because it has a chirality flip on the fermion line which cannot
 occur until the dynamical mass
has been generated ; i.e. well below the $\Lambda_{\chi}$ scale. Coming
from the     short-distance QCD
side $(\mu>\Lambda_{\chi})$ this operator simply does not exist, and
therefore $    g-2$ is of order
$1/N_c$ in agreement with the claim in ref.[4].

\vskip 0.75in

E. de R. and S.P. are very grateful to Hans Bijnens; and Roberto Peccei and
Volodya Miransky, respectively; for interesting conversations and
discussions.

\bigskip

\noindent{\bf REFERENCES}
\medskip
\parindent=1truecm
\item{\hbox to\parindent{\enskip [1]}\hfill}A. De R\'ujula, H. Georgi
 and S. Glashow, Phys. Rev.
{\bf D12} (1975) 147.

\item{\hbox to\parindent{\enskip [2]}\hfill}A. Manohar and H. Georgi,
Nucl. Phys. {\bf B234} (1984)  189.

\item{\hbox to\parindent{\enskip [3]}\hfill}G. 't Hooft, Nucl. Phys.
{\bf B72} (1974) 461.

\item{\hbox to\parindent{\enskip [4]}\hfill}S. Weinberg, Phys. Rev.
Lett. {\bf 65} (1990) 1181.

\item{\hbox to\parindent{\enskip [5]}\hfill}M. Gell-Mann and M.
 L\'evy, Nuovo Cimento {\bf 16}  (1960) 705.

\item{\hbox to\parindent{\enskip [6a]}\hfill}S. Peris,
Phys. Lett. {\bf B268} (1991) 415.

\item{\hbox to\parindent{\enskip [6b]}\hfill}S. Peris,
Phys. Rev. {\bf D46} (1992) 1202.

\item{\hbox to\parindent{\enskip [7a]}\hfill}S. Weinberg,
Phys. Rev. Lett. {\bf    67} (1991) 3473.

\item{\hbox to\parindent{\enskip [7b]}\hfill}D.A. Dicus, D. Minic,
U. van Kolck     and R. Vega, Phys. Lett. {\bf B284} (1992) 384.

\item{\hbox to\parindent{\enskip [8a]}\hfill}
S.L. Adler, Phys. Rev. Lett. {\bf 14} (1965) 1051.

\item{\hbox to\parindent{\enskip [8b]}\hfill}W.I. Weisberger, Phys.
 Rev. Lett. {  \bf 14} (1965) 1047.

\item{\hbox to\parindent{\enskip [9]}\hfill}J. Bijnens, Ch. Bruno
and E. de Rafael, Nucl. Phys. {\bf B390} (1993) 501.

\item{\hbox to\parindent{\enskip [10]}\hfill}Y. Nambu and G.
Jona-Lasinio, Phys.     Rev. {\bf 122}
(1961) 345; {\bf 124} (1961) 246.

\item{\hbox to\parindent{\enskip [11]}\hfill}D. Espriu, E. de Rafael
and J. Taron, Nucl. Phys. {\bf
B345} (1990) 22, erratum ibid. {\bf B355} (1991) 278.

\item{\hbox to\parindent{\enskip [12]}\hfill}S. Coleman and E. Witten,
Phys. Rev. Lett. {\bf 45} (1980) 100, see especially assumption \# 4
of this paper.

\break

\noindent{\bf FIGURE CAPTIONS}

Fig. 1. Tree-diagram contribution to the isospin-odd $\pi-q$ amplitude
$T^{(-)}(\nu)$. Dashed lines are pions. Full lines are constituent quarks.

Fig. 2. Set of one-loop diagrams contributing to $T^{(-)}(\nu)$, relevant
to our discussion.

Fig. 3. One-loop contribution of the $\sigma$-particle (double line) to the
amplitude $T^{(-)}(\nu)$. The dot-dashed line signifies the existence of an
imaginary part.

Fig. 4a. Bubble diagram contribution to $g_A$ from the operators in
eqs. (10). The cross at the top stands for the pion field.

Fig. 4b. Diagrammatic representation of the Schwinger-Dyson equation for
the constituent quark propagator.

\end